# Influence and Betweenness in Flow Models of Complex Network Systems

Olexandr Polishchuk

Laboratory of Modeling and Optimization of Complex Systems,
Pidstryhach Institute for Applied Problems of Mechanics and Mathematics,
National Academy of Sciences of Ukraine,
Lviv, Ukraine
od_polishchuk@ukr.net

**Abstract.** This paper provides the analysis for functional approaches of complex network systems research. In order to study the behavior of these systems the flow adjacency matrices were introduced. The concepts of strength, power, domain and diameter of influence of complex network nodes are analyzed for the purpose of determining their importance in the systems structure. The notions of measure, power, domain and diameter of betweenness of network nodes and edges are introduced to identify their significance in the operation process of network systems. These indicators quantitatively express the contribution of the corresponding element for the motion of flows in the system and determine the losses that are expected in the case of blocking this node or edge or targeted attack on it. Similar notions of influence and betweenness are introduced to determine the functional importance of separate subsystems of network system and the system as a whole. Examples of practical use of the obtained results during investigation of real complex network systems are given.

## 1. Introduction

To study any real network system (NS), whether natural or artificial, we have to form full and comprehensive representation of this system. Usually it is reached through observations, experimental and theoretical investigations and displaying the system as the models of different types [1]. When talking about network systems modeling, two main approaches may be distinguished: structural and functional. In modern NSs studies, the structural approach prevails, which is implemented in so-called theory of complex networks (TCN) [2, 3]. The subject of TCN investigations is the creation of universal network structures models, determination of statistical features that characterize their behavior and forecasting networks behavior in case their structural properties change. Sometimes the term "complex network" (CN) is used to denote both structure and system [4, 5], though these are fundamentally different concepts. The laws according to which the systems operate are usually much more complicated than the features of system structure, and methods of structural studies often do not allow us to solve NS functional problems [6]. Within the scope of functional approach, system structure is analyzed in conjunction with functions implemented by components of this structure and system in general, but the function takes precedence over structure.

The theory of binary networks is completely abstracted from the functional features of the NS. Weighted networks are an attempt to "tied" the functional characteristics of the system to the elements of structure [7]. Indeed, in each particular case, the weight of CNs edges is a reflection of certain functionality of the corresponding system [8]. Network, as a structure, is considered to be dynamic if the composition of its nodes and edges changes over time. The system is a dynamic formation, even if its structure remains unchanged. The system forms its structure in the process of development. The structure is being developed and improved from the needs of the system and not vice versa. What prompts the structure to develop, modify, or degrade? Movement of flows is one of the defining features of real NS. In some cases, providing the movement of flows is the main goal of creation and operation of such systems (transport and telecommunication systems, resource supply systems, trade and information networks, etc.), in others – the nesessary condition that provides their vital activity (blood and lymph flows, neuronal impulses in the human body). Stopping of flows movement leads to the termination of the NS existence.

Complexity of network structures and systems as well as of their models in general may be represented by different concepts. Network complexity is determined, in particular, by the presence of a large number of nodes and edges between them [9, 10]. Networks with relatively small number of elements are usually not considered complex. But these relatively small structures can generate unquestionably complex systems [11]. In other words, the complexity of the network structure is quantitative, and the complexity of the system is qualitative. While trying to embrace functional complexity we often have to neglect the structural complexity. Among the examples of this situation are the attempts to solve problems associated with controllability and observability of NSs. At the present stage, such problems are being solved for the simplest linear models of network systems with the number of nodes up to 100 [12, 13]. Such structures are hard to be called complex. At the same time, problems associated with controllability, observability and synchronization of large-scale systems are rather complex functional, not structural problems. This does not downplay the significance of structural approach of studies, as long as poor operation of many real systems is driven by the disadvantages of their structure [14, 15].

This means the need to develop a conceptual apparatus and toolkit for studying the functional features of operation process of network systems components, beginning with their elements and ending with the system as a whole. Introduction and research of functional analogues of well-known structural characteristics of complex networks elements is one of the ways to solve this problem. This allows us to compare the advantages and disadvantages of functional and structural approaches to the study of NS of different types and nature, to combine them in order to create a holistic view about the state and operation process of the system, and also contribute a deeper understanding of NS behavior and solution of some practically important problems [16, 17].

## 2. Flow Adjacency Matrices of Network Systems

The network structure is completely determined by its adjacency matrix $\mathbf{A} = \{a_{ij}\}_{i,j=1}^{N}$, where $N$ is the number of CN nodes. For the most studied binary networks, the value of $a_{ij}$ is equal to 1, if there is a connection between the nodes $n_i$ and $n_j$, and is equal to 0, if such connection is absent. Using the matrix $\mathbf{A}$ are defined the local and global characteristics of CN and studied its properties. We describe the process of system functioning on the basis of flows motion analysis by the network and introduce the following adjacency matrices of NS [16]:

1) the matrix of the density of flows which are moving by the network edges at the current moment of time $t$:

$$\boldsymbol{\rho}(t,x) = \{\rho_{ij}(t,x)\}_{i,j=1}^{N}, \quad x \in (n_i, n_j),$$

where $(n_i, n_j)$ is the edge connected network nodes $n_i$ and $n_j$, $i, j = \overline{1, N}$, $t > 0$;

2) the matrix of volumes of flows that are moving by the network edges at time $t$:

$$\mathbf{v}(t) = \{v_{ij}(t)\}_{i,j=1}^{N}, \quad v_{ij}(t) = \int\limits_{(n_i, n_j)} \rho_{ij}(t,x)dl, \quad t > 0;$$

3) the integral flow adjacency matrix of volumes of flows passed through the network edges for the period $[t-T,t]$ to the current moment $t$:

$$\mathbf{V}(t) = \{V_{ij}(t)\}_{i,j=1}^{N}, \quad V_{ij}(t) = \widetilde{V}_{ij}(t) / \max_{m,l=\overline{1,N}} \widetilde{V}_{ml}(t), \quad \widetilde{V}_{ij}(t) = \int_{t-T}^{t} v_{ij}(\tau)d\tau, \quad t \geq T > 0;$$

4) the matrix of loading of network edges at time $t$:

$$\mathbf{u}(t) = \{u_{ij}(t)\}_{i,j=1}^{N}, \quad u_{ij}(t) = v_{ij}(t) / v_{ij}^{\max},$$

where $v_{ij}^{\max}$ is bandwidth of the edge connected the network nodes $n_i$ and $n_j$, $i,j = \overline{1,N}$, $t > 0$;

5) the integral matrices of NS loading for period $[t-T,t]$ to the moment $t$:

$$\mathbf{U}^{C}(t) = \{U_{ij}^{C}(t)\}_{i,j=1}^{N}, \quad U_{ij}^{C}(t) = \max_{\tau \in [t-T,t]} u_{ij}(\tau),$$

and

$$\mathbf{U}^{L}(t) = \{U_{ij}^{L}(t)\}_{i,j=1}^{N}, \quad U_{ij}^{L}(t) = (\int_{t-T}^{t} u_{ij}^{2}(\tau)d\tau)^{1/2} / T, \quad t \geq T > 0.$$

The introduced above flow adjacency matrices in aggregate give a sufficiently clear quantitative picture of the system's operation process, allow us to analyze the features and predict the behavior of this process, to evaluate its effectiveness and prevent existing or potential threats [11, 14]. The matrices $\boldsymbol{\rho}(t,x)$ and $\mathbf{v}(t)$ can be useful for the current analysis of network system's operation. The matrix $\mathbf{V}(t)$ enable to track the integral volumes of flows that pass through the network edges. They are especially important in predicting and/or planning the NS operation and allow us to timely respond to deploying threatening processes in the system. The matrices $\mathbf{u}(t)$ and $\mathbf{U}(t)$ enable to analyze the current and integral activity or passivity of separate system components, as well as the level of their critical loading, which can lead to crashes in the NS operation. These matrices allow us to timely increase the bandwidth of network elements, build new ones or search the alternative paths of flows movement, etc. Many systems, e. g. transmission, processing and analysis of information are very dynamic formations [18]. Therefore, continuous monitoring of flows motion by the network is especially important in such systems [19]. The introduced above flow adjacency matrices allow us to carry out such real-time monitoring.

During investigation of the system and forming its model we are interested in a clear identification of the NS structure. The network elements that are not involved in the system operation will be called fictitious. Examples of the existence of numerous fictitious nodes and edges can be found in many real systems, including social networks and the Internet [10, 18]. The World Wide Web has a deep and dark web, pages of which are not indexed by any search engines [20]. Elements that are involved in the operation of particular system, but not included in its structure, will be called hidden. The identification of hidden nodes and edges plays no less important role in constructing the NS model than the search of fictitious elements. Obviously, the removal of fictitious elements contributes to overcoming the complexity problem by reducing the dimensionality of system model, and the inclusion of hidden nodes and edges – to better understanding of processes that occur in it. The flow adjacency matrices of the NS enable to identify the fictitious elements in the source network and exclude them from the system structure [11]. These matrices also allow us to carry out the search and inclusion of hidden nodes and edges in the system structure [16].

Different ways to determine both the local and global importance of the network node there are in TCN [3, 7, 21, 22]. However, the importance of a node in the structure is often not the same as the functional significance of node in the system [14].

## 3. Influence of Network Systems Nodes

The functional importance of the edge $(n_i, n_j)$ in the system is determined by the value $V_{ij}(t)$, $i, j = \overline{1, N}$. We will define the functional importance of node in the following way [16]. Denote by $v_k^{out}(t, n_i, n_j)$ the volume of flows generated in node $n_i$ and received at node $n_j$, which passed through the path $p_k(n_i, n_j)$ for the period $[t-T, t]$, $K_{ij}$ is the number of all possible paths that connect nodes $n_i$ and $n_j$, $k = \overline{1, K_{ij}}$, $i, j = \overline{1, N}$. Then

$$V^{out}(t, n_i, n_j) = \sum_{k=1}^{K_{ij}} v_k^{out}(t, n_i, n_j)$$

is the total volume of flows generated in node $n_i$ and directed to accept in node $n_j$ by all possible paths for the period $[t-T, t]$. Parameter $V^{out}(t, n_i, n_j)$ defines the strength of influence of node $n_i$ on node $n_j$ at the current time $t$, $i, j = \overline{1, N}$. Denote by $R_i^{out}(t) = \{j_{i_1}, ..., j_{i_{L_i(t)}}\}$ the set of node numbers that are the final receivers of flows generated in the node $n_i$ (Fig.1). Parameter

$$\xi_i^{out}(t) = \sum_{j \in R_i^{out}(t)} V^{out}(t, n_i, n_j) / s(\mathbf{V}(t)), \quad \xi_i^{out}(t) \in [0,1],$$

determines the strength of influence of node $n_i$ on the system as a whole, $i = \overline{1, N}$. Here $s(\mathbf{V}(t))$ is a sum of elements of the matrix $\mathbf{V}(t)$ and determines the total volume of flows which passed through the network for the period $[t-T, t]$.

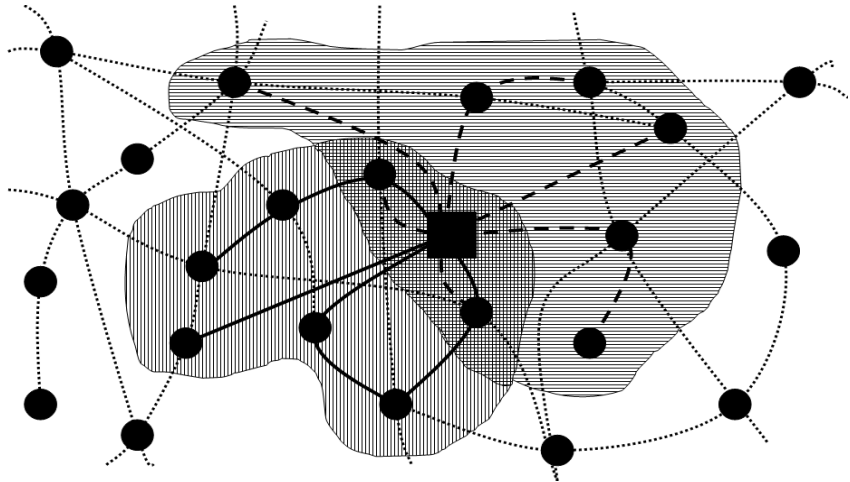

Fig. 1. Domains of input ($G_i^{in}(t)$ – vertical lines) and output influence ($R_i^{out}(t)$ – horizontal lines) of reflected by the square the node of network system.

The power of influence of node $n_i$ on the system is determined by the parameter

$$p_i^{out}(t) = L_i(t) / N, \ p_i^{out} \in [0,1],$$

where $L_i(t)$ is the number of elements of the set $R_i^{out}(t)$ which we call the domain of influence of node $n_i$ on the NS, $i = \overline{1, N}$. Denote by $\delta_i^{out}(t)$ the diameter of domain $R_i^{out}(t)$, as subnet of the source network, and $D$ – diameter of CN. Parameter

$$\Delta_i^{out}(t) = \delta_i^{out}(t) / D$$

will be called diameter of influence of the node $n_i$ on NS, $i = \overline{1, N}$. For example, the domain and diameter of influence of local government or regional media are usually limited to the relevant region of the country. At the same time, the diameter of influence of the state government and national media is equal to the "diameter" of the state as a network. The diameter of influence allows us to determine the influence of separate political parties, civic organizations, religious denominations, etc. Parameters $\xi_i^{out}(t)$, $R_i^{out}(t)$, $p_i^{out}(t)$, and $\Delta_i^{out}(t)$ will be called the output parameters of influence of the node $n_i$ respectively, $i = \overline{1, N}$. In the simplest case, the output domain of influence of each NS's node is limited by adjacent nodes. Then the power of output influence of the node is equal to its output degree, and the diameter of influence is equal to 1. In the most complex case, the output domain of influence of all NS's nodes form a complete graph. Then the power of its output influence is equal to $N$, and the diameter of influence is equal to $D$.

So-called botnets are often presented in social online services [23]. By means of these botnets one person can create the illusion of common opinion of many people, massively distribute the disinformation, organize *DDoS*-attacks, and so on. So, in one of the most popular social networks Twitter there are huge networks of fake accounts, the number of nodes of which exceeds 350 thousand [24]. Detection of nodes-generators of such botnets and their blocking allows us to prevent many negative social and economic phenomena. Parameters $\xi_i^{out}(t)$ and $p_i^{out}(t)$, $i = \overline{1, N}$, enable to identify the botnet generators with sufficient precision, since the strength and power of their influence on the NS are usually much higher than average.

The output parameters of influence of the node allow us to determine the tendencies of growth or decrease of the magnitude and power of this influence, as well as the rate and direction of its spread or convolution. Indeed, if function $\dfrac{d\xi_i^{out}(t)}{dt}$ is positive, then the strength of node's influence on the network over the period of time $[t-T, t]$ increases. If this function is negative, then this strength decreases. If function $\dfrac{dp_i^{out}(t)}{dt}$ is positive, then the power of influence of the node $n_i$ on NS increases. If the values of the function $\dfrac{d\Delta_i^{out}(t)}{dt}$ are close to 0, then the increase of number of nodes - final receivers of flows occurs in the domain bounded by the boundary $R_i^{out}(t)$. If the value of the function $\dfrac{d\Delta_i^{out}(t)}{dt}$ is positive, the diameter of the influence of the node $n_i$ increases. In general, if the values of functions $\dfrac{d\xi_i^{out}(t)}{dt}$, $\dfrac{dp_i^{out}(t)}{dt}$, and $\dfrac{d\Delta_i^{out}(t)}{dt}$ are positive, then such model adequately describes the process of spreading epidemics or computer viruses that are "generated" by one source. At the same time, the greater are the values of these functions, the faster and more threatening is this process. We note that a sharp increase of domain, power and diameter of influence is characteristic for so-called cascading failures in the network [25]. If the function $\dfrac{d\xi_i^{out}(t)}{dt}$ is negative, then the strength of node's influence on NS decreases. If functions $\dfrac{dp_i^{out}(t)}{dt}$ and $\dfrac{d\Delta_i^{out}(t)}{dt}$ are also negative then accordingly decreases the number of nodes - final receivers of flows generated in node $n_i$ as well as the power and diameter of influence of this node on the network. Thus, the output influence parameters allow us to track the dynamics of change of importance of the node $n_i$ in NS and to simulate some important processes in this system, $i = \overline{1, N}$.

Analysis of the behavior of derivatives of influence parameters allows us to determine the current trends in the state of system elements. However, the construction of at least short-term forecasts for the development of such trends is no less important [26]. Consider the algorithm for short-term forecasting of the parameter of output strength of NS's node for the period $[0, T]$. Let us the set

$$\{\xi_i^{out}(t_j)\}_{j=1}^{J}, \; J \geq 2,$$

determines the prehistory of values of this parameter at the moments of time

$$t_j = \frac{jT}{J} \in [0, T], \; j = \overline{1, J}.$$

Denote by $\boldsymbol{\Phi}(t) = \{\varphi_j(t)\}_{j=1}^{J}$ the system of linearly independent functions defined on the interval $[0, T]$. Construct a function

$$\xi_i^{out}(t) = < \mathbf{a}, \boldsymbol{\Phi}(t) >_{R^J},$$

where $\mathbf{a} = \{a_j\}_{j=1}^{J}$ is the vector of unknown coefficients. Then the forecasted value of parameter of the output strength of influence $\xi_i^{out}(t)$ of node $n_i$ on the network system at the time $t_{J+l}$ is obtained from the ratio

$$\xi_i^{out}(t_{J+l}) = < \mathbf{a}, \boldsymbol{\Phi}(t_{J+l}) >_{R^J}, l = 1, 2, \ldots, ,$$

in which vector $\mathbf{a}$ is determined from the condition

$$< \mathbf{a}, \boldsymbol{\Phi}(t_k) >_{R^J} = \xi_i^{out}(t_k), \; k = \overline{1, J}.$$

The choice of the system of basic functions can be determined by the experimentally defined behavior of the parameter of strength of influence. Similarly, short-term forecasts of the behavior for other output influence parameters of the node $n_i$, $i = \overline{1, N}$, are carried out. For the construction of medium- and long-term forecasts of the behavior of system elements, other prognostic techniques are commonly used, for example, the methods of time series [27]. However, it should be borne in mind that constructing reliable long-term forecasts of many processes occurring in real systems is often practically impossible. This is confirmed by the numerous social disturbances that have taken place in Ukraine, North Africa and the Middle East over the last decades. In most cases, it was impossible to predict the appearance of such disturbances and their magnitude even several hours before they began. Long-term forecasts of financial processes, climatic phenomena and so on are often unreliable.

Denote by $v_k^{in}(t, n_j, n_i)$ the volume of flows generated in node $n_j$ and received at node $n_i$, which passed through the path $p_k(n_j, n_i)$ for the period $[t-T, t]$, $K_{ji}$ is the number of all possible paths that connect nodes $n_j$ and $n_i$, $k = \overline{1, K_{ji}}$, $i, j = \overline{1, N}$. Then

$$V^{in}(t,n_j,n_i) = \sum_{k=1}^{K_{ij}} v_k^{in}(t,n_j,n_i)$$

is the total volume of flows generated in node $n_j$ and directed to accept in node $n_i$ by all possible paths for the period $[t-T,t]$. Parameter $V^{in}(t,n_j,n_i)$ defines the strength of influence of node $n_j$ on node $n_i$ at the current time $t$, $i,j = \overline{1,N}$. Denote by $G_i^{in}(t) = \{j_{i_1},...,j_{i_{M_i(t)}}\}$ the set of node numbers in which the flows are generated, which are sent for receiving in the node $n_i$. Parameter

$$\xi_i^{in}(t) = \sum_{j \in G_i^{in}(t)} V^{in}(t,n_j,n_i)/s(\mathbf{V}(t)), \quad \xi_i^{in}(t) \in [0,1],$$

determines the strength of influence of NS on the node $n_i$, $t \geq T > 0$, $i = \overline{1,N}$.

The power of influence of the system on the node $n_i$ is determined by the parameter

$$p_i^{in}(t) = M_i(t)/N, \ p_i^{in} \in [0,1],$$

where $M_i(t)$ is the number of elements of the set $G_i^{in}(t)$ which we call the domain of influence of NS on the node $n_i$, $i = \overline{1,N}$. Denote by $\delta_i^{in}(t)$ the diameter of domain $G_i^{in}(t)$. Parameter

$$\Delta_i^{in}(t) = \delta_i^{in}(t)/D$$

will be called diameter of influence of NS on the node $n_i$. Parameters $\xi_i^{in}(t)$, $G_i^{in}(t)$, $p_i^{in}(t)$, , and $\Delta_i^{in}(t)$ will be called the input parameters of influence of NS on the node $n_i$ respectively. In the simplest case, the intput domain of influence of each NS's node is limited by adjacent nodes. Then the power of input influence of the node is equal to its intput degree, and the diameter of influence is equal to 1. In the most complex case, the input domain of influence of all NS's nodes form a complete graph. Then the power of its intput influence is equal to $N$, and the diameter of influence is equal to $D$.

The intput parameters of influence of the node allow us to determine the tendencies of growth or decrease of the magnitude and power of this influence, as well as the rate and direction of its spread or convolution. In social networks, parameters $\xi_i^{in}(t)$ and $p_i^{in}(t)$, $i = \overline{1,N}$, allow us to identify users whose judgments pose the greatest attention of the Internet community, since the response to them (the strength and power of influence from the NS) is significantly higher than average.

Input and output parameters of influence are global dynamic characteristics of node in the NS. But determining the set of nodes-receivers of flows for a given NS's node-generator and vise versa is often an ambiguous problem. This is usually due to the type of NS and the level of flows ordering in it (for most systems with a fully ordered motion of flows – industrial, commercial, transport systems etc., the influence parameters of their nodes are sufficiently determined and predicted [14]). However, for systems with partially ordered and disordered motion of flows, the set of nodes-receivers for most or all nodes-generators and vise versa is not predetermined [11]. It should also be borne in mind that in reality the processes occured in such system and behavior of the influence parameters of the NS's nodes may be much more complicated. So a node that has directed the flow to all adjacent nodes can again become a receiver, and adjacent nodes from receivers turn into generators that direct this flow further (the spread of epidemics of infectious diseases under the so-called SIS scenario [28]). In addition, the influence parameters of NS's nodes generally are dynamic characteristics, the values of which can change significantly over time.

Special attention in TCN is given to the issue of network stability, as its ability to resist targeted external influences (hacker or terrorist attacks, etc.) [29, 30]. Attacks on the nodes with large values of input and output parameters of the strength of influence can significantly destabilize the whole system or a large part of it. These parameters allow us to define the following scenarios of attacks on the network system:

1) a list of network nodes is being prepared in order of decreasing the values of their influence strength and the nodes from the beginning of this list are consistently withdrawn from the structure until a predetermined level of critical losses is reached;

2) after removing the next node, the list of nodes formed in the previous scenario is rewritten according to the same principle and the attack is carried out on the first node from the modified list.

The second scenario takes into account the need to replace blocked nodes-generators and nodes-receivers and the corresponding redistribution of flows motion through the network. Depending on the method of dealing with potential threats, the last two scenarios can be formed separately for nodes-generators (for example, search for initiators of *DDoS*-attacks), and nodes-receivers of flows (finding the most likely targets of *DDoS*-attacks).

However, there is another side of the protection problem. It consists in the timely detection and blocking of those network system nodes that present a potential or real threat and can destabilize the system operation – hacker and terrorist groups, sources of the spread of dangerous infectious diseases, and so on. The input and output influence parameters of NS's nodes allow us to identify the botnet generators with sufficient precision. Usually, the botnet generator, by sending commands to the bots created by it (information about the purpose and content of the attack), does not need and receive no feedback, that is, for such formations, inequality

$$\frac{\xi_i^{in}(t)}{\xi_i^{out}(t)} << 1$$

is performed. From these considerations it also follows that the domain and power of output influence of such nodes are sufficiently large and the domain and power of input influence are small, moreover

$$R_i^{out} \bigcap G_i^{in} \approx 0 \,.$$

In real network systems there are practically no nodes that are only generators or receivers of flows. Indeed, the manufacturing of certain products requires the supply of raw materials and components, mining can not be carried out without the appropriate mining equipment, etc. Denote by $RG_i(t)$ the union of domains of the input and output influence of the node $n_i$, i.e.

$$RG_i(t) = R_i^{out}(t) \bigcup G_i^{in}(t) \,.$$

The interaction strength of the node $n_i$ with NS will be determined by the parameter

$$\xi_i(t) = (\xi_i^{in}(t) + \xi_i^{out}(t))/2 \,,$$

and the power of this interaction – by means of the parameter $p_i(t)$, which is equal to the number of elements of the set $RG_i(t)$.

The other side of systems resistance is its sensitivity to small changes in the structure or operation process. Such changes can be caused by both internal and external factors, and can lead to the no less consequences than targeted attacks. In this case, the stability of structure is determined by the sensitivity to small changes in the set of its nodes and edges. The structure is unstable when such changes can lead to loss of certain network properties, such as connectivity. The stability of NS operation process is determined by its sensitivity to small changes in the volume of flows motion. For example, the systems operation may become unstable in the conditions of critical loading of part of its edges (the corresponding elements of matrices $\mathbf{u}(t)$ or $\mathbf{U}_C(t)$ and $\mathbf{U}_L(t)$ are close to 1) or some the most important nodes in terms of strength and power of influence. Many systems are sensitive to small violations of established schedule of flows motion. Obviously, the stability of process is associated with the resistance of NS structure. If small changes (blocking some network nodes and edges) lead to loss of connectivity, this directly affects on the systems operation. If the load of certain elements of structure by flows is critical (close or equal to their bandwidth), it also creates a threat of blocking these elements.

Node $n_i$, for which

$$\xi_i^{in}(t) = \xi_i^{out}(t) = 0$$

and

$$W_i^{in}(t) = W_i^{out}(t) = W_i^{tr}(t) \neq 0, \; t \geq T > 0, \; i = \overline{1, N},$$

will be called a transit node. The importance of transit node in the system is determined by the volume of flows that pass through it. Extraction from structure the transit nodes is one way to reduce the dimensionality of system model. It should be borne in mind that destabilization of important transit node operation with large value $W_i^{tr}(t)$ and high betweenness centrality can destabilize the whole system or large part of it [31].

The preferential influence $\psi_i(t)$ of node $n_i$ for non-transit NS nodes we will determined by the ratio

$$\psi_i(t) = (\xi_i^{in}(t) - \xi_i^{out}(t)) / (\xi_i^{in}(t) + \xi_i^{out}(t)), \psi_i \in [-1, 1].$$

If the value of parameter $\psi_i(t)$ is close to $-1$, then the preferential influence is from the node $n_i$ on NS. If the value of parameter $\psi_i(t)$ is close to 1, then the preferential influence is from NS on the node $n_i$. In case $\psi_i(t) \approx 0$, $i = \overline{1, N}$, the influence is uniform on each side. The network structure (Fig. 2a) is usually much simpler than the structure of flows in it (Fig. 2b). Parameter of preferential influence allows us to determine the predominant direction of flows within the system (Fig. 2c).

Thus, passenger traffic in a country or a large city is characterized by the value of $\psi_i(t) \approx 0$, $i = \overline{1, N}$. At the same time, migration processes (refugee movement, urbanization, etc.) are characterized by a pronounced uneven distribution of the values of preferential influence.

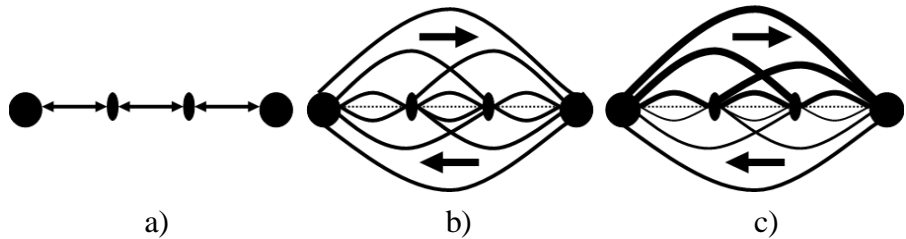

a)                    b)                    c)

Fig. 2. Fragments: a) network structure; b) structure of flows in network; c) volumes of flows motion in network.

## 4. Betweenness in Network Systems

One of the main concepts of TCN is the so-called centrality of the node, which allows us to determine its importance in the network: the most influential persons in social networks, key nodes on the Internet and transport networks, etc. [21, 32]. However, the notion "importance" may have different content, which led to the emergence of many definitions of the term "centrality". The most used measures of centrality in a complex network include degree centrality [33], closeness centrality [34], betweenness centrality [35], eigenvector centrality [36], percolation centrality [37], cross-clique [38], Katz [39], and Page Rank centralities, harmonic [40], Freeman, and alpha centralities [33] etc. At the same time, one measure of centrality may contradict another and the centrality that is important for one problem may be insignificant for another. This phenomenon was confirmed by D. Krackhardt [41], who gave an example of simple network, for which the degree, betweenness, and closeness centralities took completely different values, that is, gave three different choices of the most important nodes in system structure. Hence it follows that the mentioned above definitions of centralities have a quite relative value. This led to the introduction, along with the concepts of centrality the associated with them indicators of influence of nodes on the network structure. The main measures of the node's influence are its accessibility and expected force [22]. The accessibility of a node is determined by the number of nodes to which we can walk from it over a specified period of time. Expected force of a node's influence is determined by the number of nodes to which we can pass through two or more steps of motion (step – the transition by one edge of the network). Obviously, the above mentioned measures of centrality and influence of the node are determined solely by the properties of structure and are the characteristics of this structure, rather than system in general.

The input and output influence parameters of a node were introduced above to determine its importance in the system. These concepts allow us to quantify the participation of separate node as a receiver or generator of flows in the process of system operation and its significance in this process. Another indicator of the importance of node interaction with NS is measure of its contribution in the transit of flows through the network. One of the most used with the degree centrality in TCN is the betweenness centrality. Perhaps the notion "betweenness " is most successful in determining the participation of NS's node in the process of joint operation and interaction of all nodes in the network or a certain part of it. Therefore, to determine the functional importance of a node or an edge in the system, we will use the term "betweenness ".

Denote by $P_{ij}^{K_{ij}} = \{p_{ij}^k\}_{k=1}^{K_{ij}}$ the set of paths that connect the nodes-generators and nodes-receivers of NS flows, and contain, as an element, the edge $(n_i, n_j)$, $i, j = \overline{1, N}$. Let us $v_{ij}^k(t)$ is the volume of flows that have passed through path $p_{ij}^k$ from the node-generator to the node-receiver, and hence by the edge $(n_i, n_j)$, for the period $[t - T, t]$. Then the value

$$V_{ij}^{K_{ij}}(t) = \sum_{k=1}^{K_{ij}} v_{ij}^k(t)$$

defines the total volume of flows that have passed through the set of paths $P_{ij}^{K_{ij}}$, and hence by the edge $(n_i, n_j)$, over the same period of time. Parameter

$$\Phi_{ij}(t) = V_{ij}^{K_{ij}}(t) / s(\mathbf{V}(t)),$$

which determines the specific weight of flows passed through the edge $(n_i, n_j)$ for period $[t - T, t]$, will be called the betweenness measure of this edge in the process of NS operation.

The set $L_{ij}$ of all NS's nodes, which lie on the paths of set $P_{ij}^{K_{ij}}$, will be called the betweenness domain, and the number $\eta_{ij}$ of these nodes – the power of betweenness of the edge $(n_i, n_j)$ (Fig. 3). Denote by $\delta_{ij}$ the diameter of betweenness domain of the edge $(n_i, n_j)$. This diameter is calculated as the diameter of the set $L_{ij}$. Parameter

$$\Delta_{ij} = \delta_{ij} / D$$

will be called the diameter of betweenness of the edge $(n_i, n_j)$, $i, j = \overline{1, N}$.

The parameters of measure, domain, power and diameter of betweenness of the edge $(n_i, n_j)$ are global characteristics of its importance in the process of NS operation, $i, j = \overline{1, N}$. They, in particular, determine how the blocking of this edge will affect on the work of domain of its betweenness, the magnitude of this domain and, as a result, the whole system.

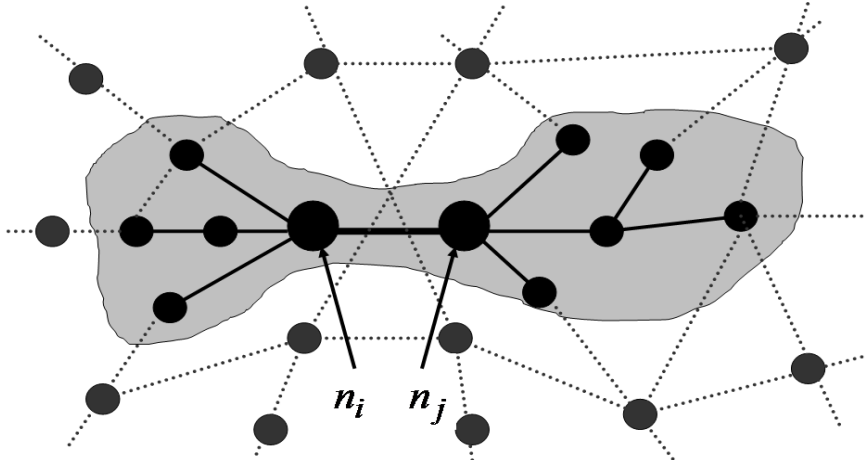

Fig. 3. The betweenness domain of edge $(n_i, n_j)$ in the process of NS operation.

Denote by $K_i$ the set of paths that connect nodes-generators and nodes-receivers of NS flows, and pass through a node $n_i$, $i = \overline{1, N}$. Let us $v_i^k(t)$ is the volume of flows passing through path $p_i^k$ from the node-generator to node-receiver, and hence through the node $n_i$, for the period $[t - T, t]$. Then the parameter

$$V_i^{K_i}(t) = \sum_{k=1}^{K_i} v_i^k(t)$$

determines the total volume of flows that have gone through the set of paths $P_i^{K_i}$, and hence through node $n_i$, over the same period of time. Parameter

$$\Phi_i(t) = V_i^{K_i}(t) / s(\mathbf{V}(t)),$$

which determines the specific weight of flows passing through the node $n_i$ for period $[t - T, t]$, will be called the betweenness measure of this node during the NS operation. The set $M_i$ of all NS's nodes, which lie on the paths of set $P_i^{K_i}$, will be called the betweenness domain, and the number $\eta_i$ of these nodes – the power of betweenness of the node $n_i$. Denote by $\delta_i$ the diameter of betweenness domain of the node $n_i$. Then parameter

$$\Delta_i = \delta_i / D$$

will be called the diameter of betweenness of the node $n_i$, $i = \overline{1, N}$.

The parameters of measure, domain, power and diameter of betweenness of the node $n_i$ are global characteristics of its importance in the process of NS operation, $i = \overline{1, N}$. They, in particular, determine how the blocking of this node will affect on the work of domain of its betweenness, the magnitude of this domain and, as a result, the whole system.

Betweenness parameters allow us to define the following scenarios of attacks on the network system:

1) a list of network nodes is being prepared in order of decreasing the values of their betweenness measure and the nodes from the beginning of this list are consistently withdrawn from the structure until a predetermined level of critical losses is reached;

2) after removing the next node, the list of nodes formed in the previous scenario is rewritten according to the same principle and the attack is carried out on the first node from the modified list.

The second scenario takes into account the need to replace blocked nodes-generators and nodes-receivers of flows and the search for alternative paths of movement of transit flows that pass through blocked nodes, i.e. the corresponding redistribution of flows motion through the network. Similar scenarios of attacks are also formed for NS's edges, since in many cases the blocking of network edge is much simpler than blocking one of the nodes that it combines. The parameters of betweenness of nodes and edges allow us to estimate to what part of the NS the consequences of failures of the corresponding system element will spread and to what losses this will result in the sense of lack of supply of certain volumes of transit flows.

We have defined above the parameters of betweenness of the node, taking into account only the transit flows that pass through it. However, the importance of betweenness parameters can be significantly expanded, taking into account that the node $n_i$ can be not only a transit, but also a generator and final receiver of flows. Then the set $P_i^{K_i}$ can be supplemented by the paths of flows that begin (generated) or end (received) in the node $n_i$. Denote such supplemented set by $\widetilde{P}_i^{K_i}$, $i = \overline{1, N}$. Then parameter

$$\widetilde{\Phi}_i(t) = (\Phi_i(t) + \xi_i^{in}(t) + \xi_i^{out}(t)) / 3$$

will be called a generalized measure of betweenness of the node $n_i$ in the process of NS operation. Accordingly, the set $\widetilde{M}_i$ of all NS's nodes, which lie on the paths from the set $\widetilde{P}_i^{K_i}$, will be called a generalized betweenness domain, and the number $\widetilde{\eta}_i$ of these nodes is the generalized betweenness power of the node $n_i$, $i = \overline{1, N}$. The generalized betweenness parameters take into account the interaction between all directly and indirectly connected nodes of NS (generators, receivers and transits) and allow us to form the most effective scenarios of attacks on them. Principles for creation such scenarios are described above.

## 5. Influence and Betweenness of Subsystems of Complex Network Systems

Denote by $S$ the subsystem of source NS, formed on the basis of principles of ordering or subordination [42]. Let us $H_S$ is the set of nodes that make up the structure of subsystem $S$, and $F_S$ is the set of edges that combine nodes of the set $H_S$.

Denote by $G_S^{out}$ the set of all nodes-generators of flows included in the set $H_S$, $p_S^{out}$ – the number of elements of $G_S^{out}$ and determine by the parameter

$$\xi_S^{out}(t) = \sum_{i \in G_S^{out}} \xi_i^{out}(t) / s(\mathbf{V}(t))$$

the strength of influence of the subsystem $S$ on NS as a whole.

Let us

$$R_S^{out} = \bigcup_{i \in G_S^{out}} R_i^{out}$$

is the set of numbers of nodes – final receivers of flows generated in nodes belonging to the set $G_S^{out}$ (Fig. 4). Divide the set $R_S^{out}$ into two subsets, namely

$$R_S^{out} = R_{S,int}^{out} \bigcup R_{S,ext}^{out},$$

where $R_{S,int}^{out}$ is a subset of nodes $R_S^{out}$ belonging to $H_S$, and $R_{S,ext}^{out}$ is a subset of nodes $R_S^{out}$ that belong to the supplement to $H_S$ in the source NS. The set $R_{S,ext}^{out}$ is called the domain of the output influence of subsystem $S$ on NS, and the number of elements $p_{S,ext}^{out}$ of this set is the power of this influence. Denote by $\delta_{S,ext}^{out}$ the diameter of the set $R_{S,ext}^{out}$. Parameter

$$\Delta_{S,ext}^{out} = \delta_{S,ext}^{out} / D$$

will be called the diameter of output influence of subsystem $S$ on NS.

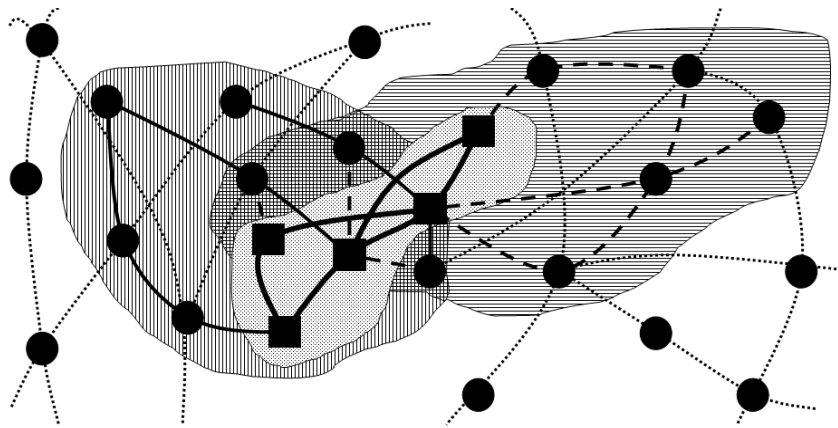

Fig. 4. Domains of output ($R_{S,ext}^{out}$ – vertical lines) and input influence ($G_{S,ext}^{in}$ – horizontal lines) of NS subsystem (subsystem nodes are reflected by squares).

The external and internal output strength of influence of the nodes-generators of flows belonging to the set $G_S^{out}$ on subnets $R_{S,ext}^{out}$ and $R_{S,int}^{out}$ will be determined by the parameters

$$\xi(t)_{S,ext}^{out} = \sum_{i \in R_{S,ext}^{out}} \xi_i^{out}(t) / s(\mathbf{V}(t))$$

and

$$\xi(t)_{S,int}^{out} = \sum_{i \in R_{S,int}^{out}} \xi_i^{out}(t) / s(\mathbf{V}(t))$$

respectively. Then the parameter

$$\omega_S^{out}(t) = \frac{\xi_{S,ext}^{out}(t)}{\xi_{S,int}^{out}(t)}$$

determines the relative strength of influence of subsystem $S$ on the system as a whole. Namely the smaller the value of parameter $\omega_S^{out}$, the less the strength of influence of the subsystem $S$ on NS. Parameters $\xi_{S,ext}^{in}$, $R_{S,ext}^{in}$, $p_{S,ext}^{in}$, $\Delta_{S,ext}^{in}$ and $\omega_S^{out}$ will be called the output influence parameters of subsystem $S$ on NS.

Denote by $R_S^{in}$ the set of all nodes-receivers of flows included in the set $H_S$ (fig. 4), $p_S^{in}$ – the number of elements of $R_S^{in}$ and determine by the parameter

$$\xi_S^{in}(t) = \sum_{i \in R_S^{in}} \xi_i^{in}(t) / s(\mathbf{V}(t))$$

the strength of influence of NS on subsystem $S$.

Let us

$$G_S^{in} = \bigcup_{i \in R_S^{in}} G_i^{in}$$

is the set of numbers of nodes-generators from which the flows are directed to nodes belonging to the set $R_S^{in}$. Divide the set $G_S^{in}$ into two subsets, namely

$$G_S^{in} = G_{S,int}^{in} \bigcup G_{S,ext}^{in},$$

where $G_{S,int}^{in}$ is a subset of nodes $G_S^{in}$ belonging to $H_S$, and $G_{S,ext}^{in}$ is a subset of nodes $G_S^{in}$ that belong to the supplement to $H_S$ in the source NS. The set $G_{S,ext}^{in}$ is called the domain of the input influence of NS on the subsystem $S$, and the number of elements $p_{S,ext}^{in}$ of this set is the power of this influence. Denote by $\delta_{S,ext}^{in}$ the diameter of the set $G_{S,ext}^{in}$. Parameter

$$\Delta_{S,ext}^{in} = \delta_{S,ext}^{in} / D$$

will be called the diameter of input influence of NS on subsystem $S$.

The external and internal input strength of influence of the nodes-receivers of flows belonging to the set $R_S^{in}$ on subnets $G_{S,ext}^{in}$ and $G_{S,int}^{in}$ will be determined by the parameters

$$\xi_{S,ext}^{in}(t) = \sum_{i \in G_{S,ext}^{out}} \xi_i^{in}(t) / s(\mathbf{V}(t))$$

and

$$\xi_{S,int}^{in}(t) = \sum_{i \in G_{S,int}^{out}} \xi_i^{in}(t) / s(\mathbf{V}(t))$$

respectively. Then the parameter

$$\omega_S^{in}(t) = \frac{\xi_{S,ext}^{in}(t)}{\xi_{S,int}^{in}(t)}$$

determines the relative strength of influence of NS on subsystem $S$. Namely the smaller the value of parameter $\omega_S^{in}$, the less the strength of influence of NS on the subsystem $S$. Parameters $\xi_{S,ext}^{in}$, $G_{S,ext}^{in}$, $p_{S,ext}^{in}$, $\Delta_{S,ext}^{in}$ and $\omega_S^{in}$ will be called the input influence parameters of NS on subsystem $S$.

The behavior of derivatives of influence parameters of NS's subsystems allows us to determine the tendencies of growth or decrease of their magnitude and power, as well as the rate of distribution and growth. For a deeper study of the behavioral patterns of these parameters, it is also advisable to use the prediction methods described above.

The notion of community is important in TCN [43]. Community is a group of closely interconnected CN's nodes which are weakly interconnected with other nodes in the network. The main disadvantage of existing methods for identifying communities in the CN (methods of minimal cut, hierarchical clusterization, modularity maximization, methods based on clicks, spectral properties of the network, evaluation of system entropy etc.) [43, 44, 45, 46] etc.) along with computational complexity and resource expenditures is the lack of reliable criterion of what the group of nodes determined by any of these methods really forms the community. A pair of parameters $(\omega_S^{out}, \omega_S^{in})$ gives such criterion. Indeed, the smaller are the values of these parameters, the less is the external interaction of subsystem $S$ with the system as a whole and the larger are intragroup interactions, which is, in essence, a community definition. Moreover, a pair of these parameters obviously allows us to determine the system-wide and internal activity or passivity of the subsystem $S$.

Determining the participation of subsystem $S$ in the system operation in the sense of predominant influence allows the parameter

$$\psi_S(t) = \frac{\xi_{S,ext}^{in}(t) - \xi_{S,ext}^{out}(t)}{\xi_{S,ext}^{in}(t) + \xi_{S,ext}^{out}(t)}, \ \psi_S \in [-1, 1].$$

If the value of parameter $\psi_S(t)$ is close to –1, then the predominant is the influence of subsystem $S$ on the NS, i. e. it is generally a subsystem that generates flows. If the value of parameter $\psi_S(t)$ is close to 1, then the influence of NS on subsystem $S$ is predominant, i. e. it is generally the receiver of flows. In the case of $\psi_S(t) \approx 0, i = \overline{1, N}$, the influence is uniform from each side, i. e. the subsystem $S$ is simultaneously both the generator of flows and the flows receiver. It is also non-difficult to determine the strength of predominant influence between two arbitrary subsystems of NS, the sets of nodes of which does not intersect.

Equally important for the analysis of NS operation are the parameters of betweenness of its separate subsystems, which we define as follows. Denote by $P_S^{K_S} = \{p_S^k\}_{k=1}^{K_S}$ the set of paths that combine the NS's nodes-generators and nodes-receivers of flows and pass through elements of the subsystem $S$. Let us $v_S^k(t)$ is the volume of flows that went through path $p_S^k$ from the node-generator to node-receiver, and hence through the elements of subsystem $S$, for the period $[t-T,t]$. Then parameter

$$V_S^{K_S}(t) = \sum_{k=1}^{K_S} v_S^k(t)$$

determines the total volume of flows that went through a set of paths $P_S^{K_S}$, and therefore through elements of the subsystem $S$, over the same period of time. Parameter

$$\Psi_S = V_S^{K_i}(t)/s(\mathbf{V}(t)),$$

which determines the specific weight of flows passing through elements of subsystem $S$ for period $[t-T,t]$, will be called the betweenness measure of this subsystem during the NS operation.

The set $M_S$ of all NS's nodes, which lie on the paths of set $P_S^{K_S}$, will be called the betweenness domain (fig. 5), and the number $\eta_S$ of these nodes – the power of betweenness of subsystem $S$. Denote by $\delta_i$ the diameter of betweenness domain of the node $n_i \in H_S$. Then parameter

$$\Delta_S = \max_{n_i \in H_S} \delta_i / D,$$

will be called the diameter of betweenness of subsystem $S$.

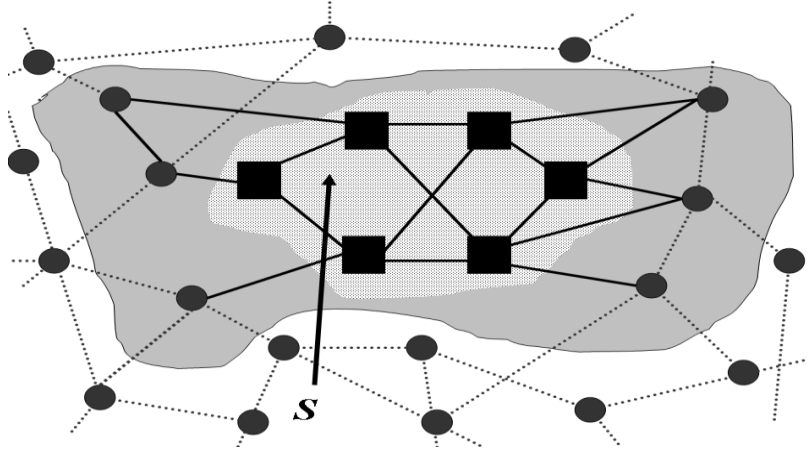

Fig. 5. Betweenness domain of subsystem $S$ in the process of NS operation.

The parameters of measure, domain, power and diameter of betweenness of subsystem $S$ are global characteristics of its importance in the process of NS operation. They, in particular, determine how the blocking of this subsystem will affect on the work of domain of its betweenness, the magnitude of this domain and, as a result, the whole system. In addition, the small values of betweenness parameters of the subsystem $S$ may also indicate that it forms a community within the NS.

The behavior of derivatives of betweenness parameters of NS's subsystems allows us to determine the tendencies of growth or decrease of their magnitude and power, as well as the rate of distribution and growth. For a deeper study of the behavioral patterns of these parameters, it is also advisable to use the prediction methods described above.

Another way to determine the most important subsystems of NS is to introduce the notion of its *k*-core, that is, the largest subnet of the source CN, all nodes of which have degree not less than *k*, and the extraction from the network structure of nodes with degree less than *k* [47]. Using the flow characteristics of NS allows us to introduce the concept of flow $\lambda$-core of network system, as the largest subnet of source network, for which all elements of the integral flow adjacency matrix have values not less than $\lambda$, $\lambda \in [0,1]$, [48]. Fig. 6a reflects the structure of railway transport system of the western region of Ukraine. The thickness of lines in this figure is proportional to the weight of edges – the volumes of flows passing through them. Fig. 6b displays the 4-core of this network and fig. 6c reflects the flow 0.7-core of this system.

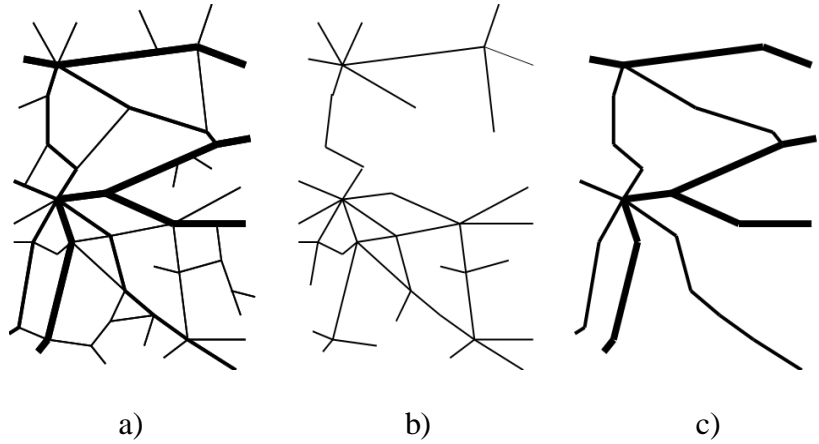

a)                    b)                    c)

Fig. 6. Fragments: a) source NS; b) 4-core of CN; c) 0.7-core of NS.

Introduce the integral flow adjacency matrix of $\lambda$-core by means of ratio

$$\mathbf{V}^{\lambda}(t) = \{V_{ij}^{\lambda}(t)\}_{i,j=1}^{N}, \quad V_{ij}^{\lambda}(t) = \begin{cases} V_{ij}(t), \text{if } V_{ij}(t) \geq \lambda \\ 0, \text{if } V_{ij}(t) < \lambda \end{cases}, \quad \lambda \in [0,1], t \geq T.$$

We will use parameter $\sigma_{\lambda}(t)$ to determine the specific weight of $\lambda$-core. This parameter is equal to the ratio of volumes of flows passing by the $\lambda$-core to the volume of flows that pass through the network as a whole during the period $[t-T,t]$:

$$\sigma_{\lambda}(t) = s(\mathbf{V}^{\lambda}(t)) \big/ s(\mathbf{V}(t)).$$

Since the main goal of the most network systems is to provide the flows motion, parameter $\sigma_{\lambda}(t)$ quantifies how the $\lambda$-core provides the implementation of this goal. Thus, this parameter determines the importance of subsystem, formed by $\lambda$-core, in the NS operation process as a whole. So, the spread of epidemics usually occurs on the ways of intensive movement of large masses of people, and the spread of computer viruses – on the paths of intense information traffic. The flow cores of NS with large values of $\lambda$ determine the most likely paths of deploying such processes.

## 6. Integral Parameters of Influence and Betweenness of Complex Network Systems

The most common indicator of NS operation is the total volume of flows that pass through the network over period of time $[t-T,t]$. This indicator is determined by the value $s(\mathbf{V}(t))$, $t \geq T$. But it is rather relative, since it does not determine how effective the system functions compared to potential opportunities.

Let us denote

$$\mathbf{V}_{\max}(t) = \{V_{ij}^{\max}(t)\}_{i,j=1}^{N}, \quad t \geq T,$$

where $V_{ij}^{\max}(t)$ is the maximum volume of flows that could pass through th edge $(n_i, n_j)$, $i,j = \overline{1,N}$, over the same time period, taking into account the bandwidth of this edge. Parameter

$$\pi(t) = \frac{s(\mathbf{V}(t))}{s(\mathbf{V}_{\max}(t))} \in [0,1], \quad t \in [t-T, t]$$

determines how effective the NS operates compared to its potential possibilities.

Critically loaded systems are very vulnerable to increasing the volume of flows. It is difficult and sometimes impossible to find alternative paths of flows motion, since such paths that can increase the volume of flows may not exist. At the same time, the most dangerous for the stable system operation is the critical loading of its $\lambda$-core with a high specific weight. This is the negative reverse side of an attempt to maximize the efficiency of NS operation, if the bandwidth of the system nodes and edges does not increase at the same time.

Denote by $G^{out}$ the set of all network nodes-generators and introduce parameter

$$V^{out}(t) = \sum_{i \in G^{out}} \sum_{j \in R_i^{out}} V^{out}(t, n_i, n_j).$$

Determine parameter

$$p^{out} = \mu(G^{out}) / N, \; p^{out} \in [0,1],$$

where $\mu(G^{out})$ is the power (number of elements) of subset $G^{out}$, which determines the specific weight of nodes-generators in the system structure. Obviously, the smaller the value $p^{out}$, the more vulnerable is the NS to destabilization the work of the nodes-generators of flows.

Denote by $R^{in}$ the set of all network nodes-receivers and introduce parameter

$$V^{in}(t) = \sum_{i \in R^{in}} \sum_{j \in G_i^{in}(t)} V^{in}(t, n_j, n_i)).$$

Determine parameter

$$p^{in} = \mu(R^{in}) / N, \; p^{in} \in [0,1],$$

where $\mu(R^{in})$ is the power of subset $R^{in}$, which determines the specific weight of nodes-receivers in the system structure. Obviously, the smaller the value $p^{in}$, the more vulnerable is the NS to destabilization the work of the nodes-receivers of flows.

Any real system is open, that is, it interacts with other systems [49]. Let us that our system is the subsystem of a bigger formation – mega-system. Then, as in the previous paragraph, we can introduce the influence parameters of our system on this mega-system and vice versa, as well as the parameters of its betweenness during mega-system operation. This approach allows us to reach the level of interdependent network system interactions research.

## 7. Conclusions

The functional approach of network systems research is considered in this article. In order th study the process of such systems operation the flow adjacency matrices of different types were introduced. It was also analyzed, how these matrices help to investigate and forecast the peculiarities of this process, evaluate its efficiency and prevent existing and potential threats. Global dynamic influence and betweenness parameters of the network systems elements were determined. These parameters allow us to identify nodes that generate and receive flows, and transit nodes, determine the predominant direction of flows within the system, study activity, passivity, and stability of separate system components and NS in general, as well as form much more realistic scenarios of potential attacks on the system, quantify the losses from these attacks, and build the more reliable means of protecting it. The parameters of influence and betweenness of network system components defined in the article, as well as the concept of its flow cores, allow us to identify the most important subsystems for NSs operation and contribute to a better understanding of the processes that occur in them. Obtained results can be used to reduce the NS vulnerability from negative external and internal influences, to develop the modern methods for information and security systems protecting, to improve the efficiency of operation of transport and industrial networks of different types, etc.